\documentclass{ws-procs9x6}
\usepackage{graphicx}

\def\etal {{\it et al.}}

\begin{document}

\title{NEW EXPERIMENTS WITH ANTIPROTONS}

\author{\vspace{-.75in}\rightline{\small\rm IIT-CAPP-10-05}\vspace{.5in}}
\author{D.M.\ KAPLAN\\ (for the AGE and P-986 Collaborations)}

\address{Physics, Illinois Institute of Technology\\
Chicago, Illinois 60616, USA\\
E-mail: kaplan@iit.edu}

\begin{abstract}
Fermilab operates the world's most intense antiproton source. Recently  
proposed experiments can use those antiprotons either parasitically 
during Tevatron Collider running or after the Tevatron Collider finishes 
in about 2011.  For example, the annihilation of 8 GeV antiprotons might 
make the world's most intense source of tagged $D^0$ mesons, and thus the 
best near-term opportunity to study charm mixing and search for new physics via its CP-violation signature.  Other possible precision measurements include properties of the $X(3872)$ and the charmonium system.  An 
experiment using a Penning trap and an atom interferometer could make the 
world's first measurement of the gravitational force on antimatter.  These 
and other potential measurements using antiprotons could yield a broad 
physics program at Fermilab in the post-Tevatron era.
\end{abstract}

\bodymatter

\section{Introduction}

\label{intro}
Several intriguing questions, some involving CPT and Lorentz symmetry violation (the themes of this Meeting), can be studied with low- and medium-energy $\overline p$ beams. These  have motivated experiments at the CERN Antiproton Decelerator\cite{AD} and the planned {$\overline {\rm P}$}ANDA experiment\cite{PANDA} at the Facility for Antiproton and Ion Research; 
as described below, such experiments are now proposed at Fermilab as well.\cite{LoI,P-986}
`Medium-energy' questions include new-physics searches  in charm mixing and  CP violation (CPV), hyperon decay, and the  $X$, $Y$, and $Z$ states, as well as antihydrogen CPT tests; at low energy, the gravitational force on antihydrogen can be measured.

Table~\ref{tab:sens-comp} compares  current and future  antiproton sources. 
The highest-intensity antiproton source is at Fermilab. Having served ${\overline p}p$ fixed-target experiments including E760 and E835, it is now solely dedicated to 
the Tevatron Collider, but could once again support dedicated antiproton experiments after completion of the Tevatron program, currently planned for  2011 (although 2014 is a possibility under discussion). 

\begin{table}[tb]
\tbl{Antiproton energies and intensities at existing and future facilities.}
{\begin{tabular}{@{}lccccc@{}}\toprule
&  & \multicolumn{2}{c}{{Stacking:}} & \multicolumn{2}{c}{{Operation:}}  \\
\raisebox{1.5ex}[0pt]{Facility} &\raisebox{1.5ex}[0pt]{ $\overline{p}$ K.E. (GeV)}& {Rate ($10^{10}$/hr)}  & D.F. & {Hrs/yr} & $10^{13}$\,$\overline p$/{yr}\\
\colrule
CERN AD& 0.005, 0.047 & -- & -- & 3800 &0.4 \\
\multicolumn{2}{@{}l}
{Fermilab Accumulator:}\\
~now& 8 & 20 & 90\% & 5550 & 100\\
~proposed &$\approx$\,3.5--8& 20 & 15\% & 5550 &17 \\
~with new ring &2--20? & 20 & 90\%& 5550 & 100 \\
FAIR ($\stackrel{>}{_\sim}\,$2018) & 2--15 &  3.5 & 90\% & 2780$^*$ & 9\\
\botrule
\end{tabular}
}
\label{tab:sens-comp}
\footnotesize
~~$^*$ The lower number of operating hours at FAIR 
arises from medium-energy antiproton operation having to share time with other programs.
\end{table}

\section{Proposed antiproton experiments at Fermilab}

\subsection{Medium-energy $
{\overline p}
$-annihilation experiment}
\label{sec:charm}
A very capable and cost-effective experiment can be mounted by adding a magnetic spectro\-meter to the E760 lead-glass calorimeter,\cite{CalNIM} using an available BESS solenoid\cite{BESS},  fine-pitch scintilla\-ting fibers (SciFi), the D\O\ SciFi readout system,\cite{D0-upgrade-NIM06} and hadron ID via fast timing.\cite{psec} This could produce world-leading measurements of charm mixing and the other effects mentioned above, provided the relevant cross sections are of the expected magnitude. 
\subsubsection{Charm mixing and  CP violation}

After a $>$20-year search, $D^0$ 
mixing is now established  at $>$\,10$\sigma$.\cite{DPF2009} While the $\approx$\,1\% mixing rate may indicate a Standard-Model origin,~\cite{Bigi-Uraltsev-Petrov} a significant  new-physics contribution (signaled by CPV) is not ruled out~\cite{Bigi09,Petrov,Grossman-etal}. Since new physics can produce differing effects in the up- and down-type quark sectors,~\cite{Bigi09,Grossman-etal}  such  studies are important not only with $s$ and $b$ hadrons, but also with charm\,---\,the only up-type meson that can mix.

Although unmeasured, somewhat above threshold ($\sqrt{s}\stackrel{>}{_\sim}$\,\,4\,GeV) many expect $\sim$\,$\mu$b $\overline{p}N$\,$\to$\,open-charm  
production.\cite{Titov,Eichten-private,PANDA-LoI} E.g., using Eq.\ (5) of Ref.\ \refcite{Braaten-X-3872}, we obtain 1.3\,$\mu$b for the $D^{*0}{\overline D}{}^{0}$ final state.
At ${\cal L}=2\times10^{32}$\,cm$^{-2}$s$^{-1}$, this is $\sim$\,$5\times10^9$ events/year. 
Target-$A$ depen\-dence~\cite{A-dep} can enhance statistics by $\sim$\,$ A^{1/3}$, giving a much larger sample than the $B$ factories' $10^9$ events. A wire or pellet target, limiting primary vertices to $\sim$\,10\,$\mu$m in $z$, can make the $D^0$ decay distance resolvable. The low charged-particle multiplicity  ($\langle n_{ch}\rangle\approx2$) at this energy may allow clean samples with the application of only modest vertex cuts, hence high efficiency. Medium-energy 
$\overline{p}N$ interactions may thus be the optimal way to search for charm CPV. 

Preliminary simulation and background studies imply 
a $D^{*\pm}\to D^0\pi^\pm$ signal-to-background ratio of $\sim$\,10-to-1 before vertex cuts. With 150\,$\mu$m resolution in  $z$, $>$\,100-to-1 signal-to-background seems possible with efficiency $\stackrel{>}{_\sim}\,$10\%. Thus we can expect to reconstruct $\sim$\,$3\times10^7$ tagged ${D}^0 \to K^-\pi^+$ events per year,  compared with some $1.2\times10^6$ events in 0.54\,fb$^{-1}$ at  Belle.~\cite{Belle}

\subsubsection{Hyperon CP violation and rare decays}

The HyperCP Experiment~\cite{E871}  detected unexpected possible signals at the $\stackrel{>}{_\sim}$\,2$\sigma$ level  for new physics in the rare decay~\cite{Park-etal} $\Sigma^+\to p\mu^+\mu^-$ and the  
$\Xi^-\to\Lambda\pi^-$ CP asymmetry:\cite{BEACH08} $A_{\Xi\Lambda}=[-6.0\pm 2.1({\rm stat}) \pm 2.0({\rm syst})] \times10^{-4}$. Since the $\overline{p}p\to\Omega^-{\overline\Omega}{}^+$ threshold lies in the same region as the open-charm threshold, the proposed  experiment  can test these observations using $\Omega^-\to\Xi^-\mu^+\mu^-$
decays and potential ${}^{^(}\overline{\Omega}{}^{^{\,)\!}}{}^\mp$ CPV (signaled by small $\Omega$--$\overline\Omega$ decay-width differences in ${}^{^(}\overline{\Lambda}{}^{^{\,)\!}}K^\mp$ or ${}^{^(}\overline{\Xi}{}^{^{\,)\!}}{}^0\pi^\mp$ final states\cite{OmegaCP}).

Extrapolation from $\overline p$$p\to \Lambda \overline \Lambda$ and $\Xi^- {\overline \Xi}{}^+$ implies $\sigma(\overline p$$p\to \Omega^- {\overline \Omega}{}^+)$\,$\approx$\,\,60\,nb just above threshold, or $\sim$\,$10^8$ events/year. What's more, the measured $\approx 1\,$mb cross section\cite{Chien-etal}  for associated hyperon production means $\sim$\,$10^{12}$ events/year, which could  directly confront the HyperCP evidence (at 2.4$\sigma$ significance) for a possible new particle of mass 214.3\,MeV/$c^2$ in the three observed $\Sigma^+\to p\mu^+\mu^-$ events.\cite{Park-etal} 
Further in the future, the dedicated $\overline{p}$ storage ring of Table~\ref{tab:sens-comp} might decelerate antiprotons to the $\Lambda\overline{\Lambda}$, $\Sigma^+\overline{\Sigma}{}^-$, and $\Xi^-\overline{\Xi}{}^+$ thresholds, for a comprehensive program testing hyperon CPV.

\subsubsection{Precision measurements in the charmonium region}

E760 and E835 made the world's most precise  ($\stackrel{<}{_\sim}$\,100\,keV) measurements of charmonium masses and widths,\cite{E760-chi,E835-psi-prime} thanks to the precisely known collision energy of the stochastically cooled $\overline p$ beam and the H$_2$-jet target. 
Significant charmonium-related questions remain, most notably the nature of the mysterious $X(3872)$ state\cite{ELQ} and improved measurements of the $h_c$ and $\eta^\prime_c$.\cite{QWG-Yellow} The width of the $X$ may well be $\ll$\,1\,MeV.\cite{Braaten-Stapleton} The unique ${\overline p}p$ precision is what is needed to establish whether the $X(3872)$ is a $D^{*0}\overline{D}{}^0$ molecule.\cite{molecule}

The ${\overline p}p\to X(3872)$  formation cross section may be similar to that of the $\chi_c$ states.\cite{ditto,Braaten-X-3872} 
The E760 $\chi_{c1}$ and $\chi_{c2}$  detection rates of 
1 event/nb$^{-1}$  at the mass peak,\cite{Armstrong-chi_c2} along with the lower limit 
${\cal B}[X(3872) \to\pi^+\pi^- J/\psi] > 0.042$ at 
90\% C.L.,\cite{BaBar-BR} imply that at the peak of the $X (3872)$, 
about 500 
events/day can be observed. Although CDF and D\O\ could also
amass  $\sim$\,$10^4$ $X (3872)$ decays, backgrounds and energy resolution 
limit 
their incisiveness.) 
Large samples will also be obtained in other modes besides $\pi^+\pi^- J/\psi$, increasing the statistics 
and improving knowledge of $X (3872)$ branching ratios.

The above may be an under- or an overestimate, perhaps by as much as an order of magnitude. 
Nevertheless, it appears that a new experiment at the Antiproton Accumulator could obtain 
the world's largest clean samples of $X (3872)$, in perhaps as little as a month of running. 
The high statistics, event cleanliness, and unique precision available in the ${\overline p}p$ formation 
technique could enable the world's smallest systematics. Such an experiment could thus 
provide a definitive test of the nature of the $X (3872)$.

\subsection{Antihydrogen experiments}
\subsubsection {In-flight {CPT} tests} 
Production of antihydrogen in flight\cite{Blanford}  may offer a way around some of the difficulties encountered in the CERN 
trapping experiments. Methods to measure the antihydrogen Lamb shift and fine structure 
have been proposed.\cite{Blanford-Lamb-shift} Progress towards this goal may be 
compatible with normal Tevatron Collider operations (a possibility currently under investigation), and the program could continue into the post-Tevatron era. 

\subsubsection{Antimatter Gravity Experiment}
While General Relativity predicts identical gravitational forces on matter and antimatter,  a direct experimental test has yet to be made.\cite{Fischler-etal} Quantum  gravity can 
include non-tensor forces that cancel for matter-matter 
interactions but add for 
matter-antimatter ones. Possible Òfifth forces,Ó non-$1/r^2$ dependence, and Lorentz violation have also been discussed.\cite{Kostelecky} The acceleration of antimatter ($\overline g$) in the earth's gravitational field is sensitive to these effects. 
Such a measurement for antihydrogen ($\overline{\rm H}$) has only recently become feasible and is now approved at the AD\cite{AEGIS} and proposed at Fermilab.\cite{LoI}
The Fermilab proposal\cite{LoI} seeks to form a  slow ($\approx$\,1\,km/s) $\overline{\rm H}$ beam in 
a Penning trap and pass it through an atom interferometer, using either material gratings (giving $\delta g/g\sim10^{-4}$) or  laser techniques\cite{Chu} ($\delta g/g\sim$\,$10^{-9}$).
Fermilab's high $\overline p$ flux  means that even an inefficient ($\sim$\,10$^{-4}$) deceleration approach gives enough antiprotons for competitive measurements. Deceleration ideas start with the Main Injector, probably useable down to $\approx$\,400\,MeV, followed by  an `antiproton refrigerator,'\cite{refrigerator} reverse linac, or  small synchrotron.\cite{LoI}

\section{Outlook}

When the Tevatron Collider program completes, new and unique measurements can be made at the Fermilab Antiproton Source.\cite{New-pbar,LEAP08} Such a program can substantially
broaden the clientele and appeal of US particle physics.

\end{document}